\documentstyle[psfig]{paper}
\newcommand{\gtsima}{$\; \buildrel > \over \sim \;$}
\newcommand{\ltsima}{$\; \buildrel < \over \sim \;$}
\newcommand{\simgt}{\lower.5ex\hbox{\gtsima}}
\newcommand{\simlt}{\lower.5ex\hbox{\ltsima}}

\newcommand{\T}{ {\scriptscriptstyle {\rm T}} }
\newcommand{\sz}{ {{\rm sz}} }
\newcommand{\COBE}{ {\scriptscriptstyle {\rm COBE}} }

\newcommand{\keV}{ {\rm keV} }
\newcommand{\mpc}{ {\rm Mpc} }
\newcommand{\dttsz}{{\hbox
      {$\displaystyle\left({\delta T \over T}\right)_\sz$} }}
\newcommand{\dtt}{{\hbox
      {$\displaystyle\left({\delta T \over T}\right)$} }}
\def\pp{\par\parshape 2 0truecm 14.0truecm 1truecm 13.0truecm\noindent}

\begin{document}

\title{
THE X-RAY HALO OF THE LOCAL GROUP AND THE CMB QUADRUPOLE
 }

\author{Yasushi SUTO \\
{\it 	Research Center for the Early Universe, 
School of Science,\\
 The University of Tokyo, Tokyo 113,  JAPAN}}

\maketitle

\section*{Abstract}

Since recent X-ray observations have revealed that most clusters of
galaxies are surrounded by an X-ray emitting gaseous halo, it is
reasonable to expect that the Local Group of galaxies has its own
X-ray halo. We show that such a halo, with temperature $\sim 1\keV$
and column density $\sim O(10^{21}) {\rm cm}^{-2}$, is a possible
source for the excess low-energy component in the X-ray background.
The halo should also generate temperature anisotropies in the
microwave background via the Sunyaev-Zel'dovich effect. Assuming an
isothermal spherical halo with the above temperature and density, the
amplitude of the induced quadrupole turns out to be comparable to the
COBE data without violating the upper limit on the $y$-parameter.

\section{Introduction}

The origin of the X-ray background (XRB) remains one of the most
challenging problems in X-ray astrophysics. Figure 1 summarizes the
current results on the XRB energy spectra, $I(\varepsilon)$ below
10\keV observed with different satellites.  As is known (McCammon \&
Sanders 1990; Fabian \& Barcons 1992), $I(\varepsilon) \sim 10
\varepsilon^{-0.4}{\rm keV}\cdot{\rm s}^{-1}\cdot{\rm
sr}^{-1}\cdot{\keV}^{-1}$ is a good empirical fit over the range 3 to
10 \keV, where $\varepsilon$ denotes X-ray energy in units of \keV.
In the soft band, however, the observed spectra do not seem to be in
perfect agreement among different satellites; both Einstein IPC (Wu et
al. 1991; plotted in diamonds) and ROSAT (Hasinger 1992; Shanks et al. 
1991; upper-left lines) data suggest a large excess below 2\keV. More
recently the Japanese X-ray satellite ASCA (crosses) reported a modest
but significant excess soft component below 1\keV relative to the
extrapolation of the above power-law fit in the higher energy band
(Gendreau et al. 1995). Thus the existence of the soft excess is well
established although its amplitude is still somewhat controversial.

\begin{figure}
\begin{center}
   \leavevmode\psfig{figure=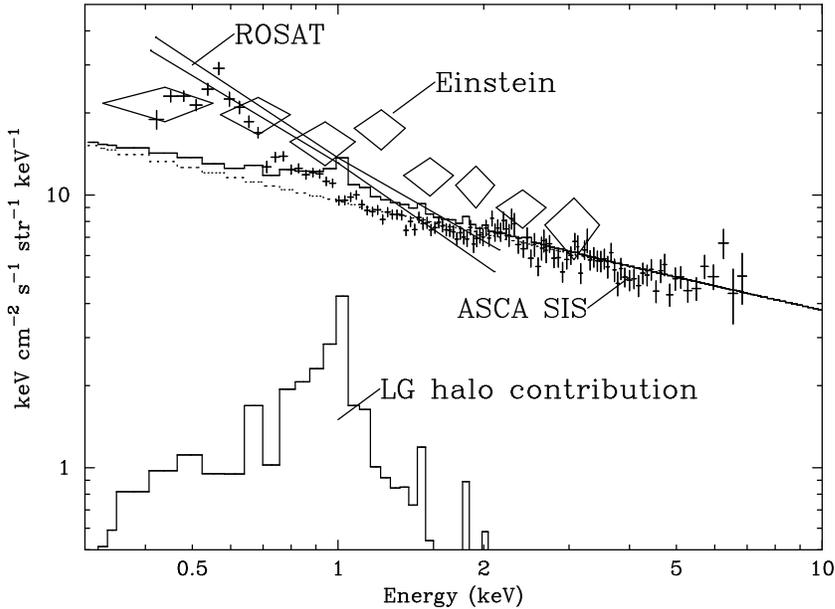,height=8cm,angle=-90}
\end{center}
\caption{XRB spectra observed with different X-ray satellites and
expected
contribution from the halo of the Local Group (Suto et al. 1996).
\label{fig:xrb}
}
\end{figure}

Since Galactic absorption becomes important in the soft X-ray energy
band, the origin of this excess component allows several possibilities
including Galactic sources, extragalactic point-like sources (Hasinger
1992; Shanks et al. 1991), a diffuse thermal component (Wang \& MaCray
1993), the accumulation of the thermal bremsstrahlung emission from
the low temperature low-density plasma surrounding distant clusters of
galaxies (Cen et al. 1995; Kitayama \& Suto 1996).  Figure 2 shows an
example of theoretical predictions of the contribution of clusters on
the XRB in the standard cold dark matter model ($\Omega_0=1$,
$\lambda_0=0$, $\Omega_b=0.06$, $h=0.5$) which is normalized by COBE
(Sugiyama 1995).  Here $\Omega_0$, $\lambda_0$, $\Omega_b$, and $h$
are the cosmological density parameter, the dimensionless cosmological
constant, the baryon density parameter, and the Hubble constant in
units of $100${\rm km/s/Mpc}.  Three lines indicate the theoretical
predictions on the basis of the Press-Schechter theory but with
different assumption on the cluster evolution (Kitayama \& Suto 1996
for details).  Also plotted are the observed data from {\it ASCA} with
statistical errors (Gendreau et al. 1995), and the simulation by Kang
et al. (1994; triangles). This suggests that the clusters of galaxies
can in fact account for the observed soft-excess in the XRB.
\begin{figure}
\begin{center}
   \leavevmode\psfig{figure=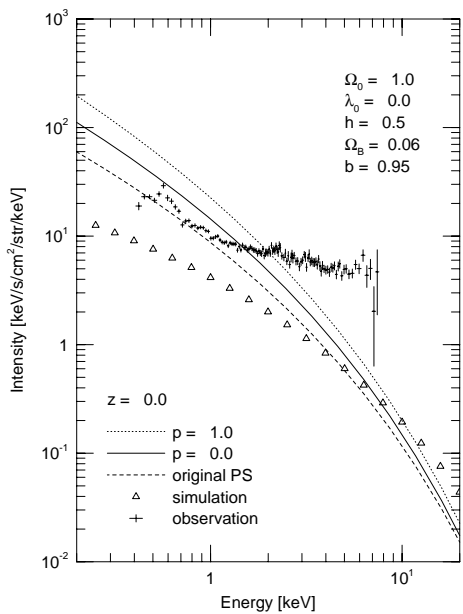,height=8cm,angle=0}
\end{center}
\caption{The XRB spectra contributed from clusters of galaxies
(Kitayama \& Suto 1996).
\label{fig:ksxrbeds}
}
\end{figure}

\section{Sunyaev-Zel'dovich effect due to the halo of the Local Group}

Yet another possibility which we propose here is the emission from an
X-ray halo of the Local Group (LG). Since the gaseous halos of
clusters of galaxies are known to be strong sources of X-rays, it is
reasonable to assume that the LG has its own X-ray halo.  To be more
specific, suppose that the LG is associated with a spherical
isothermal plasma whose electron number density is given by
\begin{equation}
\label{eq:ner}
n_e(r) = n_0 {r_c^2 \over r^2 + r_c^2},
\end{equation}
where $n_0$ is the central density and $r_c$ is the core radius.  If
we are located at distance $x_0$ off the LG center (Fig. 3), the
electron column density at angular separation $\theta$ from the
direction to the center is
\begin{equation}
\label{eq:ne}
N_e(\mu) = \int_0^{\infty} { n_0 r_c^2 d\xi \over 
\xi^2 -2x_0\mu \xi + x_0^2 + r_c^2 } 
= {n_0 r_c^2 \over x_0}{1 \over \sqrt{a^2 - \mu^2}}
\left[{\pi \over 2}+\sin^{-1}\left({\mu \over a}\right)\right] ,
\end{equation}
where $\mu \equiv \cos\theta$, and $a\equiv \sqrt{1+(r_c/x_0)^2}$.  

In Fig. 1 the LG halo contribution to the XRB spectra simulated with
the Raymond-Smith model assuming an isothermal plasma temperature
$T=1\keV$, $r_c=0.15\mpc$, $x_0=0$, $n_0 = 10^{-4} {\rm cm}^{-3}$ and
0.3 times solar abundances is also plotted(lower-left histogram).
This LG halo model has a total X-ray luminosity of $10^{41} {\rm
erg}\cdot{\rm s}^{-1}$ between $0.5 \keV$ and $4\keV$, and an electron
column density of $6\times10^{20} {\rm cm}^{-2}$.  If we add this
component to $I(\varepsilon)= 9.6 \varepsilon^{-0.4}{\rm keV}\cdot{\rm
s}^{-1}\cdot{\rm sr}^{-1}\cdot{\keV}^{-1}$, the soft excess around
1\keV is explained (solid line).  Note that it is not our primary
purpose here to find the best fit parameters because the single
power-law component is simply an extrapolation from the higher energy
band and also because the possible Galactic absorption is not taken
into account here.  Nevertheless it is interesting to see that the LG
halo can be a possible origin for the soft excess.

\begin{figure}
\begin{center}
   \leavevmode\psfig{figure=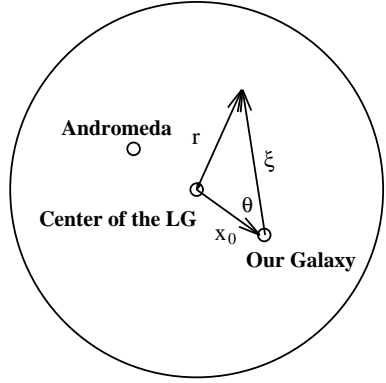,height=5cm,angle=-90}
\end{center}
\caption{Geometry of the X-ray halo of the Local Group.
\label{fig:lghalo}}
\end{figure}

If the X-ray halo of the LG really exists and accounts for the soft
excess component in the XRB, it should also produce temperature
anisotropies in the cosmic microwave background (CMB) via the
Sunyaev-Zel'dovich (SZ) effect (Zel'dovich \& Sunyaev 1969; Cole \&
Kaiser 1988; Makino \& Suto 1993; Persi et al. 1995).  In the
Rayleigh-Jeans regime, the SZ temperature decrement is given by
\begin{equation}
\label{eq:dttsz}
\dttsz(\mu) = -2 {kT \over mc^2}\sigma_\T N_e(\mu),
\end{equation}
where $k$ is the Boltzmann constant, $m$ is the electron mass, $c$ is
the velocity of light, and $\sigma_T$ is the cross section of the
Thomson scattering. We compute the multipoles expanded in spherical
harmonics:
\begin{equation}
\label{eq:dttexpand}
\dtt(\theta,\varphi) = 
\sum_{l=0}^\infty \sum_{m=-l}^{l} a_l^m \, Y_l^m(\theta,\varphi).
\end{equation}
With eqs.(\ref{eq:ne}) and (\ref{eq:dttexpand}), one obtains
\begin{equation}
\label{eq:al0}
(a_l^0)_\sz = -2 \sqrt{(2l+1)\pi}{kT \over mc^2}\sigma_\T
\int_{-1}^{1} N_e(\mu) P_l(\mu)\, d\mu ,
\end{equation}
where $P_l(\mu)$ are the Legendre polynomials. Averaging over the sky,
the above SZ anisotropies are expected to contribute in quadrature to
the CMB anisotropies as
\begin{equation}
\langle \dtt^2 \rangle = { 1\over 4\pi} \sum_{l=0}^\infty
(2l+1) \langle |a_l^m|^2 \rangle 
= { 1\over 4\pi} \sum_{l=0}^\infty (a_l^0)_\sz^2 
\equiv \sum_{l=0}^\infty (T_{l,\sz})^2 .
\end{equation}
The corresponding monopole, dipole and quadrupole anisotropies reduce
to
\begin{eqnarray}
\label{eq:t0}
T_{0,\sz} &=& \pi \theta_c 
     \,\sigma_\T {kT \over mc^2} {n_0 r_c^2 \over x_0} ,\\
\label{eq:t1}
T_{1,\sz} &=& 2\sqrt{3} \left( 1 - {r_c \over x_0} \theta_c\right) 
     \,\sigma_\T {kT \over mc^2} {n_0 r_c^2 \over x_0} ,\\
\label{eq:t2}
T_{2,\sz} &=& {\sqrt{5}\pi \over 4}
\left[\theta_c - 3{r_c \over x_0} + 3\left({r_c \over x_0}\right)^2
\theta_c \right] 
     \,\sigma_\T {kT \over mc^2} {n_0 r_c^2 \over x_0} ,
\end{eqnarray}
where $\theta_c \equiv \tan^{-1}(x_0/r_c)$.  

The {\sl COBE} FIRAS data (Mather et al. 1994) imply that the Compton
y-parameter, $y$, should be less than $2.5\times10^{-5}$ (95\%
confidence level).  With eq.(\ref{eq:t0}), this upper limit is
translated to
\begin{equation}
\label{eq:n0limit}
n_0 r_c^2/x_0 < 1.1\times10^{22} 
 \left({1.17 \over \theta_c}\right)
 \left({y \over 2.5\times10^{-5}}\right)
 \left({1 \keV \over T}\right) \, {\rm cm}^{-2}.
\end{equation}
For example taking $r_c=0.15\mpc$ and $x_0=0.35\mpc$, the constraint
(\ref{eq:n0limit}) indicates that
\begin{eqnarray}
\label{eq:t1limit}
T_{1,\sz} &<& 3\times10^{-5} \left({y \over 2.5\times10^{-5}}\right) ,\\
\label{eq:t2limit}
T_{2,\sz} &<& 1.3\times10^{-5} \left({y \over 2.5\times10^{-5}}\right) .
\end{eqnarray}
The analysis of the 1st 2 years' {\sl COBE} DMR data (Bennett et
al. 1994; Wright et al. 1994), on the other hand, yields $T_{1,\COBE}
= (1.23\pm0.09)\times10^{-3}$, and $T_{2,\COBE} =
(2.2\pm1.1)\times10^{-6}$. Therefore the LG X-ray halo can potentially
have significant effect on the quadrupole of the CMB anisotropies,
while its effect on dipole is totally negligible compared to the
peculiar velocity of the LG with respect to the CMB rest frame.

\section{Implications for the CMB temperature anisotropies}

Primordial density fluctuations with power spectrum $P(k) \propto k^n$
induce CMB anisotropies via the Sachs-Wolfe effect with multipoles
(e.g., Peebles 1993)
\begin{equation}
\label{eq:swcl}
C_l \equiv \langle |a_l^m|^2 \rangle \propto
{\Gamma(l+n/2-1/2) \over \Gamma(l-n/2+5/2)} ,
\end{equation}
where $\Gamma(\nu)$ is the Gamma function.  Therefore the
standard Harrison-Zel'dovich ($n=1$) spectrum predicts that
\begin{equation}
{l(l+1)C_l \over 6C_2} = 1 .
\end{equation}
The analytic expressions for the higher multipoles (eq.[\ref{eq:al0}])
are quite complicated. Instead we have numerically computed $(C_l)_\sz
\equiv (a_l^0)_\sz^2 /(2l+1)$ which are plotted in Fig. 4.
\begin{figure}
\begin{center}
   \leavevmode\psfig{figure=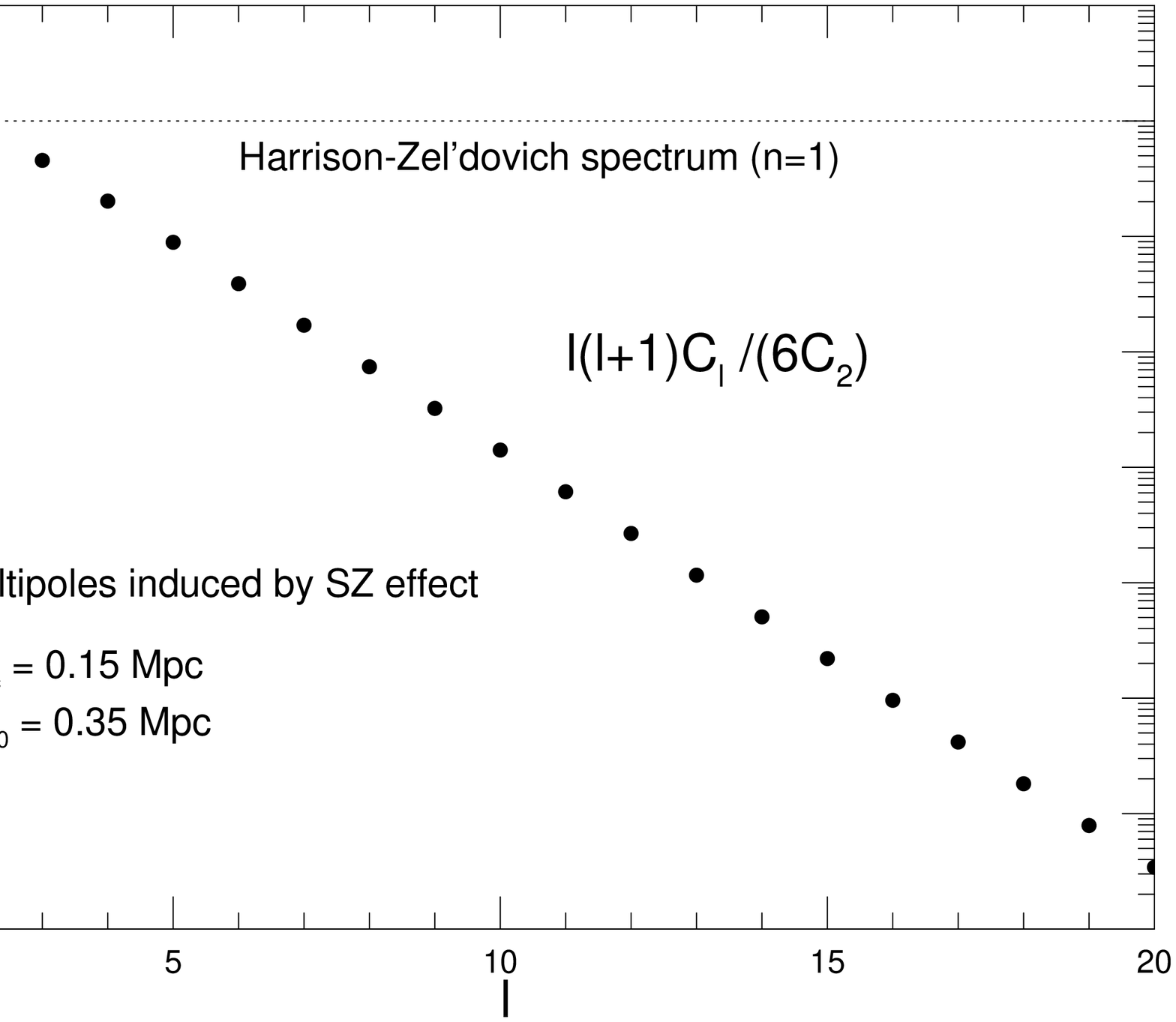,height=7cm,angle=180}
\end{center}
\caption{ Multipoles induced by the Sunyaev-Zel'dovich effect of the
Local Group halo (Suto et al. 1996). The primordial
Harrison-Zel'dovich spectrum prediction $l(l+1)C_l/ (6C_2)=1$ is
plotted in dotted line.
\label{fig:szcl} 
}
\end{figure}
Clearly the higher multipoles decrease very rapidly with $l$. Thus
even if the CMB quadrupole were contaminated by the SZ effect
described here, the higher moments would be relatively free from such
an effect and can be interpreted to reflect the true cosmological
signature (the octapole may be affected to some extent). 

Although the contribution of a distant cluster of galaxies to the
multipoles is small, its cumulative effect over the high-redshift may
be observable in the small-scale CMB anisotropies (Bennett et
al. 1993; Makino \& Suto 1993; Persi et al. 1995).  Figure 5 shows
such examples in cold dark matter (CDM) and primeval isocurvature
baryon (PIB; Peebles 1987; Suginohara \& Suto 1992) models together
with the observational upper limit from the triple beam switching
experiment at Owens Valley Radio Observatory (Readhead et al. 1989).
The temperature anisotropies predicted in CDM models with $n=1$ are
below the current observational limit, while PIB models are compatible
with the limit only for the $n=-1$ models with $\Omega_0 \simlt 0.1$,
$\lambda_0=1-\Omega_0$, and $h=0.5$ (Makino \& Suto 1993 for details).
The $n=0$ PIB models have too much small-scale power leading to the
larger SZ temperature fluctuations than the $n=-1$ models.

\begin{figure}
\begin{center}
   \leavevmode\psfig{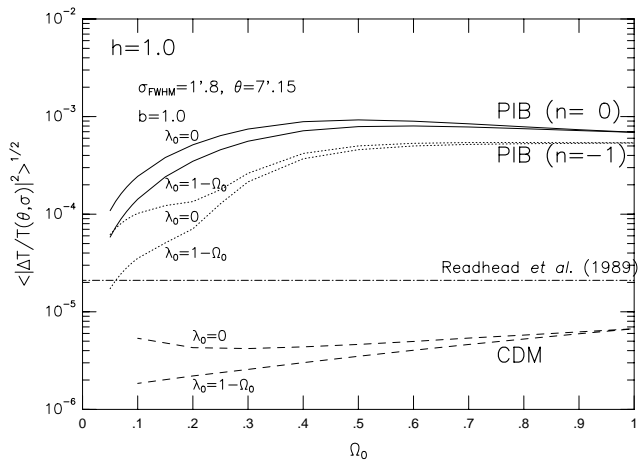}
\end{center}
\caption{ Small-scale temperature anisotropies in the CMB due to the
SZ effect of distant clusters of galaxies
(Makino \& Suto 1993).
\label{fig:szdtt} 
}
\end{figure}

Incidentally it is interesting to note that the {\it rms} quadrupole
amplitude from the {\sl COBE} 2 years' data, $Q_{\rm rms} =
(6\pm3)\mu$K, is significantly smaller than that expected from the
higher multipoles (Bennett et al. 1994; Wright et al. 1994); if one
fixes $n=1$, for instance, the power spectrum fitting using
eq.(\ref{eq:swcl}) requires that the most likely amplitude should be
$Q_{\rm rms-PS} = (18.2\pm1.5) \mu$K.  It is somewhat common to
ascribe the difference to cosmic variance. It is possible, however, to
account for it in terms of our model described here, depending on the
actual pattern of the primordial temperature fluctuations.

\section{Discussion}

The above argument can be used in putting constraints on the
properties of the possible LG halo from the COBE data. As summarized
in Fig. 6, however, the parameter range which is required for the LG
halo to provide the excess soft component is largely consistent with
the current COBE data. In addition, the LG X-ray halo should produce a
dipole signature (toward M31 and the opposite direction) in the soft
excess component; the flux $f$ plotted in Fig. 1 corresponds to what
should be observed at the center of the halo. If we adopt
$x_0=0.35$Mpc, the flux towards M31 should be $1.27f$ while $0.73f$
for the opposite direction.  Such a level of XRB variation is
detectable with careful data analysis of, for instance, the ASCA GIS
observation.

\begin{figure}
\begin{center}
   \leavevmode\psfig{figure=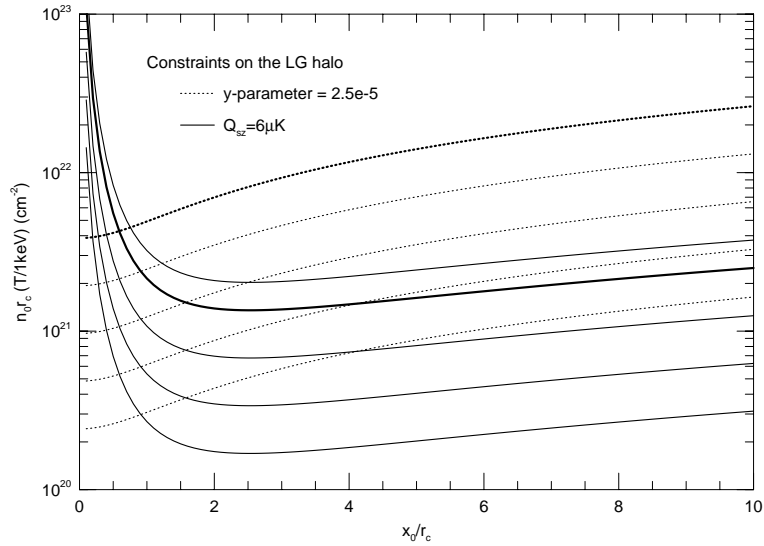,height=7cm,angle=180}
\end{center}
\caption{ Constraints on the density and the size of the halo of the
Local Group from CMB anisotropies (Suto et al. 1996). Dotted curves
correspond to $y=2.5\times10^{-5}$ (thick), $2.5\times10^{-5}/2$,
$2.5\times10^{-5}/4$, $2.5\times10^{-5}/8$, and $2.5\times10^{-5}/16$.
Solid curves correspond to $Q_{\rm sz}=9\mu$K , $6\mu$K (thick) ,
$3\mu$K , $1.5\mu$K , and $0.75\mu$K.
\label{fig:szn0}
 }
\end{figure}

Our model described above assumes a fairly idealistic density profile
(\ref{eq:ner}). A more realistic profile of the halo including
non-sphericity and spatial inhomogeneity in temperature and density
will have a stronger effect on the higher multipoles
($l\geq3$). Therefore one might even probe the properties of the LG
halo through the multipoles of the CMB map.  The direct X-ray
detection of, or constraints on, the LG halo component is of great
importance in deriving the primordial spectral index $n$ and the
amplitude of the density fluctuations from the {\sl COBE} data.

\bigskip
\bigskip

The present work is based on my collaboration with Yoshitaka Ishisaki,
Tetsu Kitayama, Nobuyoshi Makino, Kazuo Makishima, and Yasushi
Ogasaka. This research was supported in part by the Grants-in-Aid by
the Ministry of Education, Science and Culture of Japan (07740183,
07CE2002).

\bigskip
\bigskip

\def\apjpap#1;#2;#3;#4; {\pp#1, {\it #2}, {\bf #3}, #4.}
\def\apjbook#1;#2;#3;#4; {\pp#1, {#2} (#3: #4)}
\def\apjppt#1;#2; {\pp#1, #2.}
\def\apjproc#1;#2;#3;#4;#5;#6; {\pp#1, {#2} #3, (#4: #5), #6.}

\apjpap ~1. Bennett,C.L. et al. 1993;ApJL;414;L77;
\apjpap ~2. Bennett,C.L. et al. 1994;ApJ;436;423;
\apjpap ~3. Cen,R., Kang,H., Ostriker,J.P., \& Ryu,D. 1995;ApJ;451;436;
\apjpap ~4. Cole,S. \& Kaiser,N. 1988;MNRAS;233;637;
\apjpap ~5. Fabian,A.C. \& Barcons,X. 1992;ARA \& A;30;543;
\apjpap ~6. Gendreau,K.C. et al 1995;Pub.Astron.Soc.Japan.;47;L5;
\apjproc ~7. Hasinger, G. 1992; The X-ray Background;
eds. Barcons, X. \& Fabian, A.C.;Cambridge University Press;
Cambridge;229;
\apjppt ~8. Kitayama,T. \& Suto,Y. 1996;MNRAS, in press;
\apjpap ~9. Makino,N. \& Suto,Y. 1993;ApJ;405;1;
\apjpap 10. Mather,J.C. et al. 1994;ApJ;420;439;
\apjpap 11. McCammon,D. \& Sanders, W.T. 1990;ARA \& A;28;657;
\apjpap 12. Peebles, P. J. E. 1987;ApJ;315;L73;
\apjbook 13. Peebles,P.J.E. 1993;Principles of Physical Cosmology;
Princeton University Press;Princeton;
\apjpap 14. Persi,F.M., Spergel,D.N., Cen,R., \& Ostriker,J.P. 1995;ApJ;442;1;
\apjpap 15. Readhead, A. C. S., Lawrence, C. R., Myers, S. T.,
Sargent, W. L. W., Hardebeck, H. E., \& Moffet, A. T. 1989; 
 ApJ;346;566;
\apjpap 16. Shanks, T. et al. 1991;Nature;353;315;
\apjpap 17. Suginohara, T. \& Suto, Y. 1992; ApJ;387;431;
\apjpap 18. Sugiyama, N. 1995; ApJS;100;281;
\apjppt 19. Suto, Y., Makishima, K., Ishisaki, Y., \& Ogasaka,
Y. 1996;ApJL, in press;
\apjpap 20. Wang,Q.D., \& MaCray,R. 1993;ApJ;409;L37;
\apjpap 21. Wright,E.L., Smoot,G.F., Bennett,C.L., \&
Lubin,P.M. 1994;ApJ;436;443;
\apjpap 22. Wu,X., Hamilton,T, Helfand,D.J., \& Wang, Q. 1991;ApJ;379;564;
\apjpap 23. Zel'dovich,Ya.B. \& Sunyaev,R.A. 1969;Ap.Sp.Sci.;4;301;

\end{document}